\documentclass[intlimits,twoside,a4paper]{article}

\usepackage{graphicx}
\usepackage[cp1251]{inputenc}

\usepackage[centertags]{amsmath}

\DeclareMathAlphabet{\mathpzc}{OT1}{pzc}{m}{it}

\newcommand{\id}{\rd}

\newcommand{\bbZ}{{\mathbb Z}}

\def\bea{\begin{eqnarray}}
\def\eea{\end{eqnarray}}
\def\ba{\begin{array}}
\def\ea{\end{array}}

\usepackage[eqsecnum]{cmpj3}

\issue{2017}{20}{3}{33002}
\doinumber{10.5488/CMP.20.33002}

\title[Activity induced first order transition for the current in a disordered medium]{Activity induced first order transition for the current in a disordered medium\footnote{This paper is dedicated to the memory of J.-P. Badiali.}}
\author[T. Demaerel, C. Maes]{T. Demaerel, C. Maes}
\address{Instituut voor Theoretische Fysica, KU Leuven, Belgium}

\date{Received May 4, 2017, in final form June 13, 2017}

\begin{document}
\maketitle
\begin{abstract}
It is well known that particles can get trapped by randomly placed obstacles when they are pushed too much.
We present a model where the current in a disordered medium dies at a large external field, but is reborn when the activity is increased. By activity we mean the time-variation of the external driving at a constant time-averaged field.   A different interpretation of the resurgence of the current is that the particles are capable of taking an infinite sequence of potential barriers via a mechanism similar to stochastic resonance.
We add a discussion regarding the role of ``shaking'' in processes of relaxation.
\pacs 05.40.-a   05.60.-k
\end{abstract}

\section{Introduction}
Caging is a widely discussed effect for diffusion and transport in disordered media \cite{snit,Barma,fre,ben,tra}.  It is also well known that currents may vanish (or die) when the external field exceeds some threshold. A typical cause indeed is that strong driving keeps the particles trapped in regions without exit in the direction of the field. Thermal fluctuations permit escape to some point, and their strength determines the field threshold. Above that threshold, the current is zero. In this paper, we give an example of a zero-temperature first order transition for the current in a random environment as a function of the activation, from zero to some finite value.  It physically realizes  what has been called a dynamical phase transition.  The activation or ``shaking'' is realized by the time-dependence in the driving which counterbalances the possible localization effects induced by disorder and inhomogeneities.

Activity is a more recent but a very active subject of statistical mechanics.  The most immediate reference is towards the many studies today of active particles and of active matter more generally.  The concept of dynamical activity has, however, also appeared in more formal constructions of nonequilibrium physics, in particular extending it beyond soft matter.   It has also appeared in the exploration of dynamical phase transitions, and in response theory, we have been speaking about the frenetic contribution \cite{bae,kolk}.  It makes contact with older concerns like in the understanding of glassy behaviour and jamming transitions.  The fact that disorder as well as interactions can slow down relaxation and can even cause localization of matter or energy is an important issue within the study of dynamical activity \cite{gar,dpt,chan,kin}.

In the present paper, we concentrate on some simple specific toy-models where the effect of activation to get a higher conductivity is particularly clear.  It is presented as a proof of a principle while various realizations in condensed matter could greatly vary.  We start in the next section~\ref{ndc} by retelling the story of how negative differential conductivity can be seen as an effect of negative correlation between current and dynamical activity. We show that the non-monotonousness of the current gets removed by activation. Section~\ref{ide} contains our main result with the first order transition of the current in a zero-temperature overdamped dynamics.  In the last section, we present some general ideas on activation and its relation to the notion of dynamical phase transition.

\section{Removing negative differential conductivity}\label{ndc}

We recall here, by way of introduction, a number of considerations that have appeared elsewhere \cite{kolk,bae,mob}. At the end of this section we can then give a first illustration of how dynamical activation can avoid the decrease of current as a function of external driving.

\begin{figure}[!b]
\centering
\includegraphics[width=11 cm]{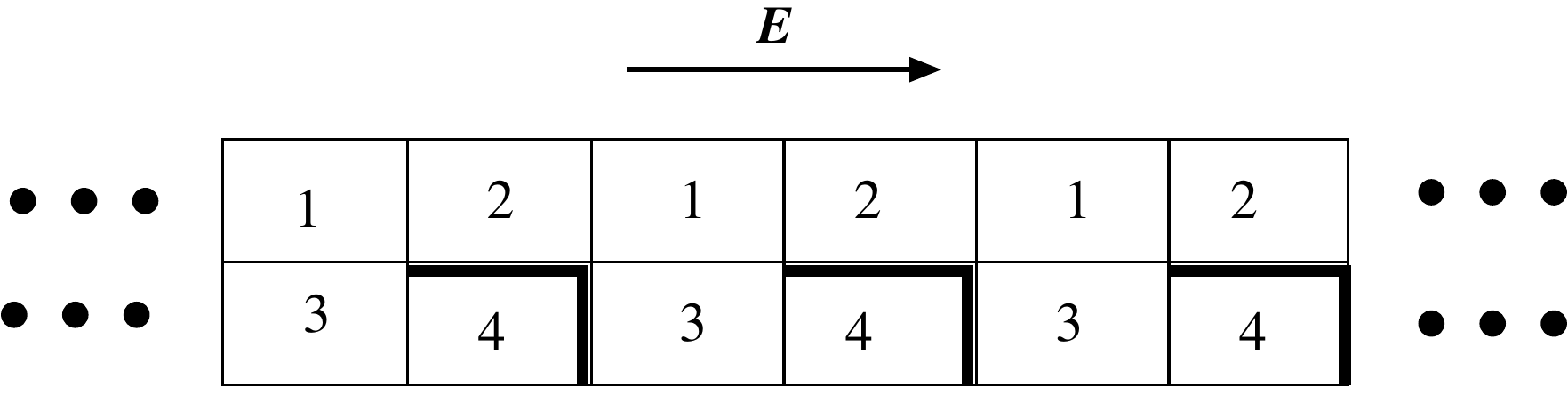}
\caption{Random walker in channel with cages; the thick dark lines are impenetrable walls in which the walker gets trapped for a time that increases with the strength of the field $S$. (Figure reproduced from~\cite{bae}.)}\label{fig:cm}
\end{figure}

We start with the mathematically trivial example of a biased random walker in continuous time to enable the emergence of a conceptually more general idea.
We put it on the one-dimensional lattice with transition rates for $x \to x+1$ given by $k(x,x+1)=p$ and similarly $k(x,x-1)=q$ for jumps to the left (where $p,q>0$ are fixed and sum to 1).  That is just a simple Markov process possibly representing a great wealth of physical situations of particles being pushed through a channel possibly as a coarse graining of a much more inhomogeneous or disordered dynamics. Physically, the two parameters $p$, $q$ are determined by an applied external field $E$ but also by more kinetic details of the channel through which the particle is moving.  A first natural correspondence is to ask that the ratio $p/q = \exp(\beta E d)$ where $\beta$ is the inverse temperature and $d$ is the distance of the hopping.  The product $s(E)=\beta E d$ is then the entropy flux (or dissipated work per temperature) per $k_{\text B}$, and we have thus assumed a form of a local detailed balance.  Yet, $p$ and $q$ are not determined completely by their ratio $p/q$.  A further physical characterization is to introduce the escape rate,
\[
p+ q =\xi(E), \qquad \frac{p}{q} = \re^{s(E)},
\]
and say how exactly it depends on the applied field $E$.  It gives the inverse of the expected residence time at each cell or position.  The stationary current $J=J(E)$, of course, depends on it:
\begin{equation}\label{effe}
J = p- q = \xi(E)\,\frac{\re^{s(E)}-1}{\re^{s(E)}+ 1}\,.
\end{equation}
In particular, when $\xi(E)$ decreases with $E$ for large $E$, then so will the current $J$ there.  We have, in other words, a simple mechanism for negative differential conductivity where $\id J/\id E <0$ for large enough $E$.  It is easy to find various physical realizations.  Typically, the decrease of $\xi(E)$ is caused by the architecture of the channel where due to some roughness, the particles can get trapped in dead ends. See, e.g., \cite{bae} for more details on a model inspired by \cite{Zia}, and represented in figure~\ref{fig:cm}.  In many cases, a current will keep flowing, however, for no matter how large driving $E$.  If, however, the traps (dead ends) have a distribution where the residence time can become infinite for $E>E_{\text c}$, then the current will  vanish for $E\geqslant E_{\text c}$. That scenario is exactly realized for certain random walks in random environments; see \cite{sol}.  A well-known example has been constructed by Barma and Dhar for a walker on a percolation cluster \cite{Barma}.   In that paper they also present a one-dimensional version which has stimulated us to find an activation recipe and which is part of the next section.

\begin{figure}[!t]
 \centering
 \includegraphics[width=10 cm]{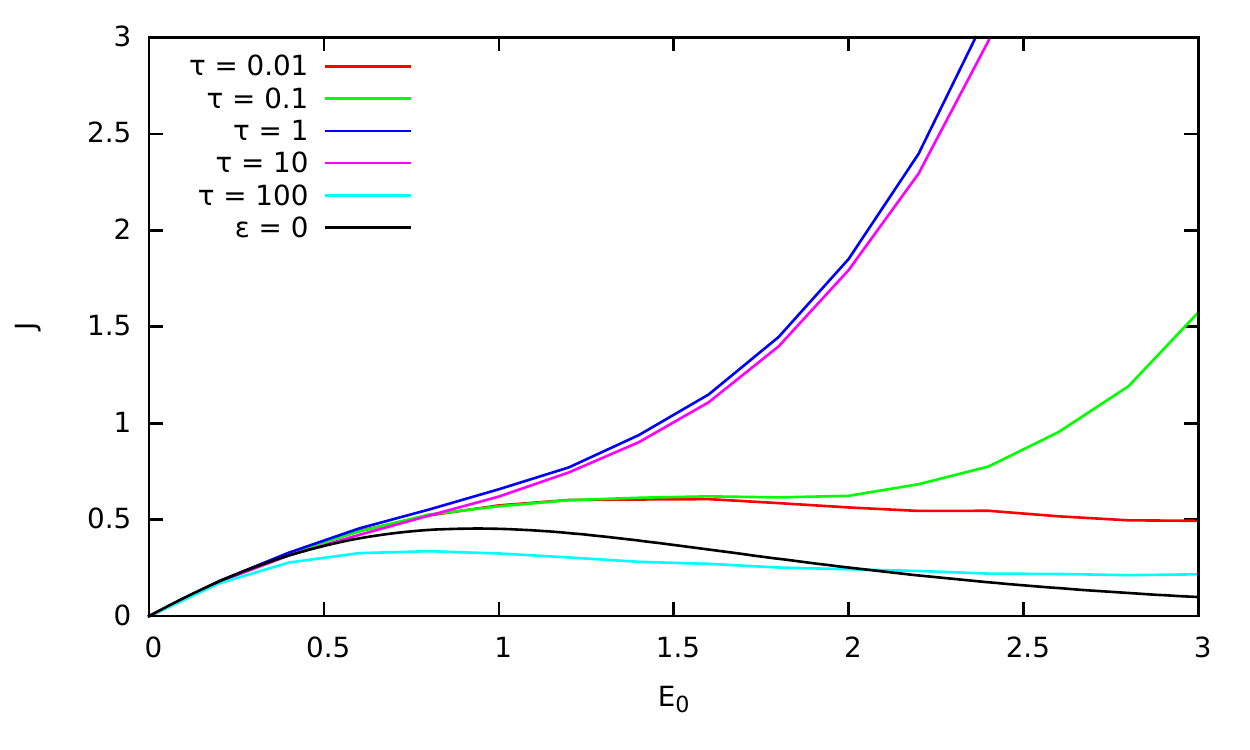}
 \caption{(Color online) The time-averaged current as a function of the driving amplitude $E_0$ for different values of the period $\tau$ of modulation. The solid black curve is when $E$ does not depend on time.}\label{dmu}
\end{figure}

Before we continue with the true transition (where the previously vanishing current resurrects by a modulated field)
we still add here a specific realization of the above where, by modulation of the driving, the negative differential conductivity disappears.
Consider the random walker as in figure~\ref{fig:cm}.  There is a long periodic channel consisting of identical cells each divided into $4$ parts, labelled $i=1,\dots, 4$. The transition rates are
\begin{equation}\label{rat}
k_\rightarrow = \re^{ E},\qquad   k_\leftarrow  = \re^{- E},\qquad\quad\;\; k_\uparrow =  1,\;\;\qquad \;\;\, k_\downarrow = 1
\end{equation}
as a function of the driving $E$.  One imagines a wall between parts $2$ and $4$ of the same cell and from $4$ to $3$ of the next cell in the forward direction. There are periodic boundary conditions in the horizontal direction.
The stationary current initially increases (in the linear response regime) but then decreases for a large field; see the solid black line in figure~\ref{dmu}.  There is a negative differential conductivity around $E=1$.  As explained in detail in \cite{bae}, the negative differential mobility can more generally be attributed to the frenetic contribution.  As an effective model, it is the escape rate $\xi(E)$ in \eqref{effe} which  decreases exponentially in $E$ which causes the non-monotonousness of the current as a function of $E$.

 Look now at figure~\ref{dmu}.  There we take the same rates as \eqref{rat} in figure~\ref{fig:cm} but where the driving field $E$ is time-dependent, following
\[
 E(t) = E_0 [1 +\epsilon\cos( \omega t)], \qquad \omega = 2 \piup / \tau
 \]
with either $\epsilon=0$ (no time-dependence) or $\epsilon=1$ (modulation). Figure~\ref{dmu} is the plot of the time-averaged current for different values of the period $\tau$.  As seen there, intermediate frequencies appear indeed best to avoid the negative differential conductivity.  The linear response regime is mostly unaffected but at large driving $E_0$, the modulation makes a serious difference, erasing the negative differential conductivity for the period $\tau$ of order 1--10.

We take  that example as an illustration of what is perhaps a more general idea; the modulation of the field is capable of liberating the particles from the traps and gives back a monotonous current as a function of the driving amplitude.  The fact that the period must be {\it tuned} is not surprising and will depend on the depth of the traps.  Another viewpoint is as follows: the walker  has to repeatedly overcome a potential hurdle whose height grows with the amplitude; modulation at the correct range of frequencies makes the particles overcome these barriers in the same sense as happens in a stochastic resonance \cite{scholar}.  We are now ready to turn that scenario in a first order dynamical phase transition in the next section.

\section{Death and resurrection of current}\label{ide}

\begin{figure}[!t]
	\centering
	\includegraphics[width=7 cm]{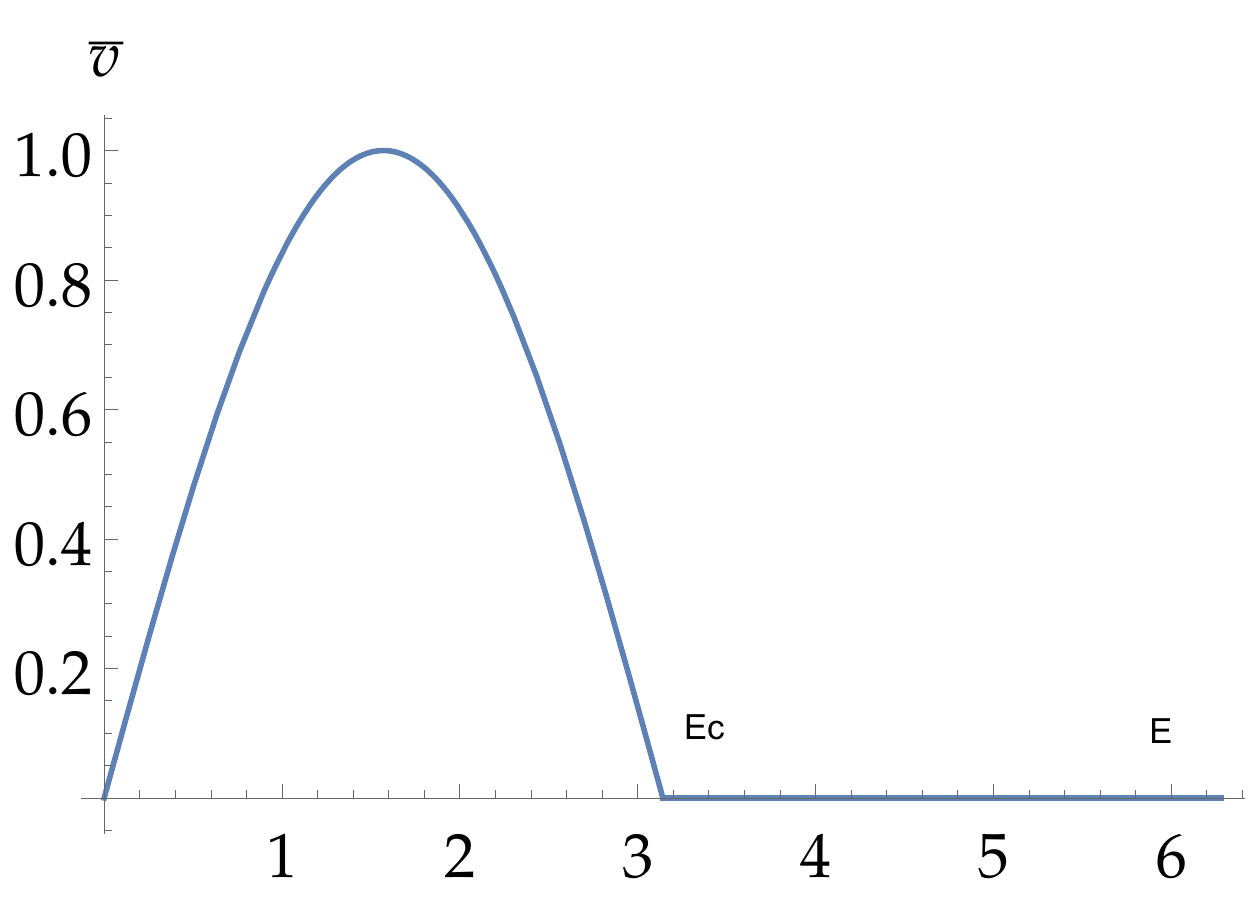}
	\caption{(Color online) A cartoon of typical current-field relation, e.g., for \eqref{EOM}, for a wire with random geometry. See also \cite{Barma}.}\label{death}
\end{figure}

That currents may die as a function of the external driving is clear from the ideas in the previous section.  The negative differential conductivity there is in essence the same phenomenon but where the scale of the trap depths is fixed. When the traps are arbitrarily deep with a not too small probability for such an exceeding depth, the dying is unavoidable.
Let us first explain this again via the random walk model but now we must be explicit about the random environment. The latter is represented via a collection $w=\{ w_x \}_{x\in \bbZ}$ of independent Bernoulli random variables, 1 with probability  $\rho$, and 0 with probability $1-\rho$ and the density $\rho \in [1/2, 1]$.  We then take the walker with rates
\[
k(x,x+1) = pw_x + (1-p)(1-w_x),\qquad k(x,x-1) = (1-p)w_x + p(1-w_x).
\]
In words, we think of $\log p/(1-p)$ as the strength of the field, but depending on the location, the particle has to move with or against that field.  At $x$ where $w_x=1$, there is a bias to the right for $p>1/2$, and there is a bias to the left at $w_x=0$.  Obviously then, there are unbounded (exponentially distributed) intervals where the field is pointing to the left, even though there is a larger density of places where the field is to the right.
We know from \cite{sol} that there is an asymptotic current,
\begin{align}
J(\rho,p) &= \frac{(2p-1)(\rho-p)}{ \rho (1-p) + p (1-\rho)} >0 \quad \text{ when } \quad  p<\rho\leqslant 1,\nonumber\\
J(\rho,p) &= 0 \quad\text{ if } \quad \frac 1{2}\leqslant \rho \leqslant p.
\end{align}
For a given density $\rho>1/2$ we will have zero current when $p$ grows larger than $\rho$.
That mathematical scenario as in figure~\ref{death} can be physically realized in various ways.  A well-known example is that of Barma and Dhar \cite{Barma} where a random walker is driven on a percolation cluster.  They also present a one-dimensional model there and our model next has been inspired by that setting.

\subsection{Random wires}
\begin{figure}[!t]
\vspace{-6mm}
	\centering
	\includegraphics[width=12 cm]{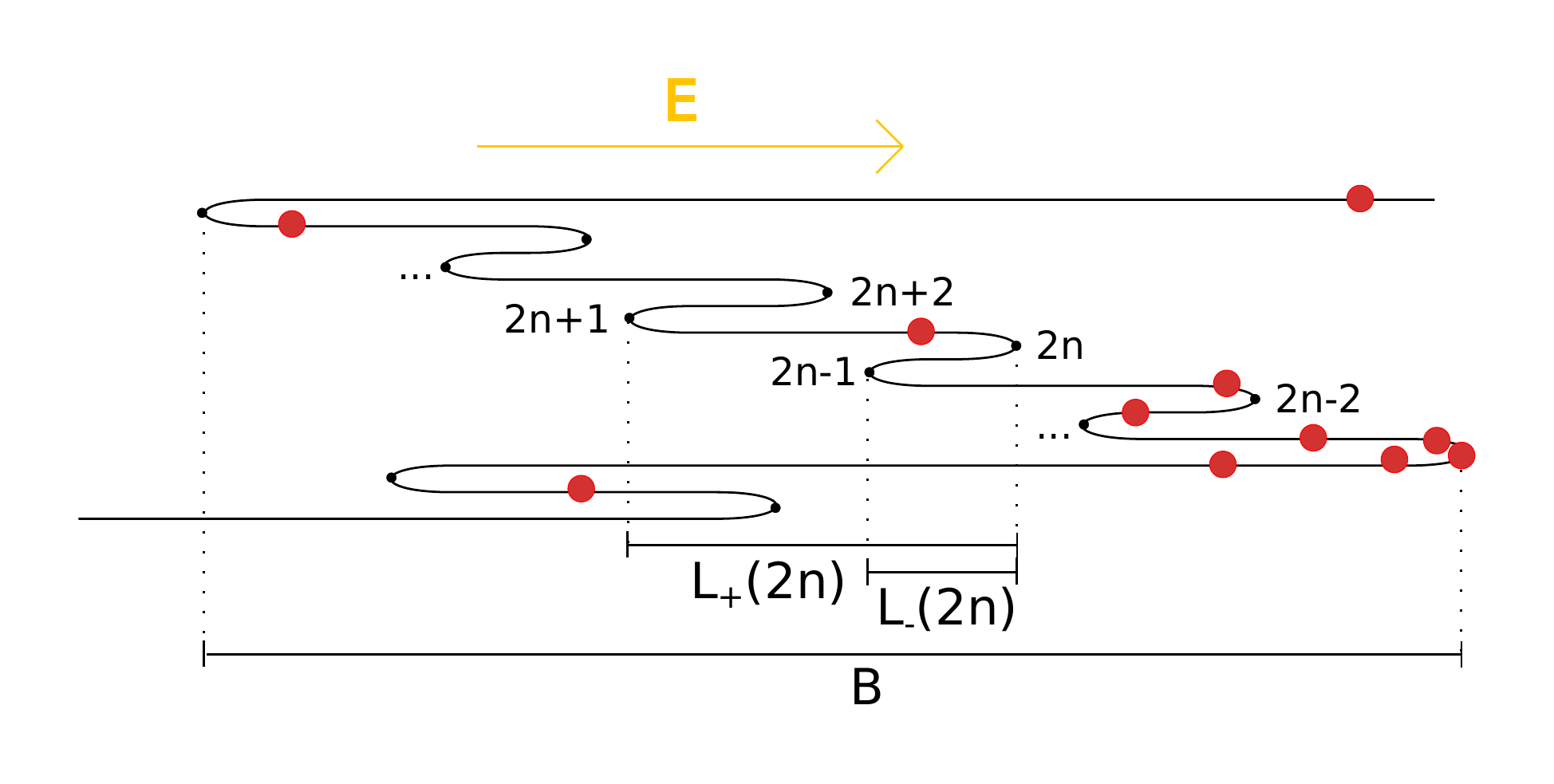}
	\caption{(Color online) The random wire.  The folds of the wire or \emph{nodes} are drawn here with a sub-infinite curvature, but they are approximated mathematically  to be single-point-folds connected with straight strands of wire. $B$ is the effective trapping length. If there are traps of large depth $B$ with density of occurrence no smaller than $C\exp(-\alpha B)$ ($C,\,\alpha>0$ fixed), then the current vanishes as soon as $E/T>\alpha$. To further clarify the setup of the model, we indicated the random strand-lengths $L_{\pm}(2n)$.} \label{trapping}
\end{figure}

We imagine an infinitely long wire which is folded at random points to give the zig-zag-like structure exhibited in figure~\ref{trapping}. We think of left to right as the overall bias of an external field and we will consider the current to the right (upward to the right in figure~\ref{trapping}). The black curve is the current-carrying wire with the red balls indicating particles, possibly responsible for a macroscopic current or a diffusion process.
Let us denote by $x\in \mathbb{R}$ an arc-length coordinate parameterizing the wire, increasing as we move along the wire in the forward direction, and let $x_n$ denote the sequence of nodes.  Between these nodes there are strands of wire but their length is random.  The even nodes at $x_{2n}$ are turning points for the wire, taking a left-to-right stretch $x_{2n-1}\rightarrow x_{2n}$ into a stretch $x_{2n}\rightarrow x_{2n+1}$ going to the left.  From every even node $2n$ there is a strand forward (towards $2n+1$) with length $L_+(2n)=x_{2n+1}-x_{2n}$ which are taken independently  and identically distributed.  Similarly, the lengths of the backward strands $L_-(2n)=x_{2n}-x_{2n-1}$ are random and identically distributed but that distribution is different from that of the $L_+$.  We typically want  the expected lengths $\mathbb{E}(L_-) > \mathbb{E}(L_+)$ to have a net trend to the right when the average external field points to the right. In fact, we suppose here that $L_+$ is bounded, and for simplicity we can assume simply that $L_+=L$, where $L$ is some finite cut-off length, which will enable the phenomena to be discussed.

A first natural dynamics on the random wire would be to take an overdamped dynamics for the position $x_t$ of the particle,
\begin{equation} \label{EOM}
\dot{x}=\bar{E}\sigma(x) + \sqrt{2T}\, \xi_t\,,
\end{equation}
where $\sigma(x)$ is $+1$ in the intervals $(x_{2n-1},x_{2n})$ and $\sigma(x) = -1$ in the intervals $(x_{2n},x_{2n+1})$ (where moving forward in the wire is moving against the bias of the field). Keeping the temperature $T$ fixed in front of the standard white noise $\xi_t$, it can be shown that while for small bias $\bar{E}$ one may see a response in the form of a small current, the random geometry of the wire may often lead to a total arrest of that current for all $\bar{E}\geqslant E_{\text c}$.

Indeed, rare but occasionally very deep traps such as in figure~\ref{trapping} may trap a particle for ever longer time-intervals, leading to a gradual stagnation. The disproportionate bias $\bar{E}$ quells the dynamic activity and, therefore, the current as in figure~\ref{death}.

A way out of this choking is to shake up the internal fabric of the wire by introducing a time-modulation in the field $E$. The frequencies that are needed depend on various scales of trapping.  However, instead of continuing with the dynamics \eqref{EOM} we go for a  model which is closer to the random walks we had in the previous section, and where a first order transition to resurrection of the current can be experienced.

\subsection{Resurrection dynamics}
The dynamics of the (non-interacting)  particles, when subjected to a strong slowly time-dependent external field $E(t)$, is taken as an overdamped zero-temperature dynamics and runs as follows:
\begin{enumerate}
\renewcommand{\labelenumi}{(\arabic{enumi})}
\item when the particle is not in a node, it travels in the same direction as the field with velocity $v(t)=E(t)$;
\item when the particle is in a node where it is not allowed to proceed in the direction of the field. It remains momentarily stuck there, just as long as this field orientation persists;
\item when the particle is in a node where the two adjacent strands are in the direction of the field, it immediately leaves the node on one of both strands. The protocol for choosing either strand is random: probability 1/2 for either option and independent of all other elements of the dynamics.
\end{enumerate}

\begin{figure}[!t]
\vspace{-4mm}
	\centering
	\includegraphics[width=8 cm]{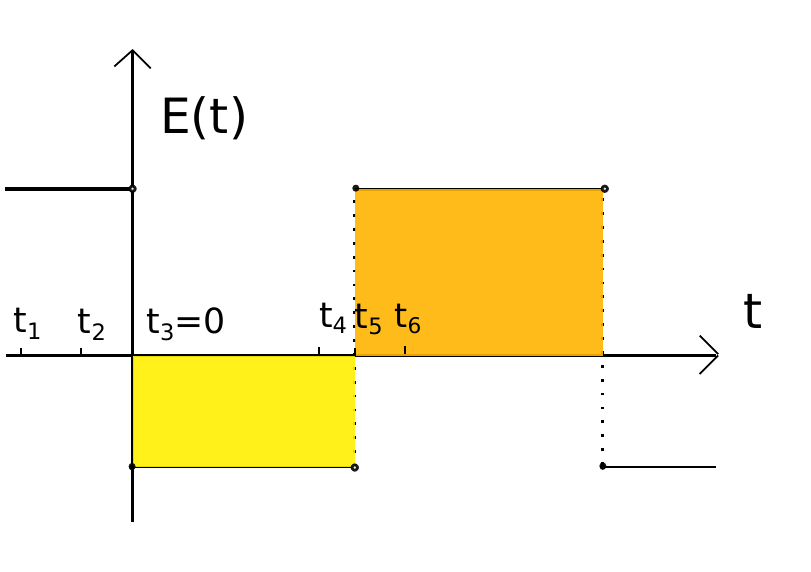}
	\caption{(Color online) The time-dependent field.  The area enclosed by its graph and the $t$-axis (shaded yellow) $\ell$ is the leftward tendency, and similarly, the area enclosed above the $x$-axis (shaded orange) $r$ is the rightward tendency.  The bias is $E_0 = (r-\ell)/\tau$ for period $\tau$. We activate the system for fixed $E_0,\tau$ by increasing $\ell$ and $r$.  The times $t_i$ refer to what happens in figures~\ref{conductivity}, \ref{conductivity3}, \ref{conductivity5}.} \label{field}
\end{figure}

The field $E(t)$ is a piecewise constant, with period $\tau$; see figure~\ref{field}.  There is a first time-interval $[0,h]$ where the field is negative equal to $-\varepsilon_1$ with product $\ell = h\,\varepsilon_1$ representing the leftward tendency of the field.  Then, in the time-interval $[h,\tau]$, the field is positive equal to $\varepsilon_2$ with right-ward tendency $r = (\tau-h)\,\varepsilon_2$.  The average bias equals $E_0 = (r-\ell)/\tau$.  When we fix $E_0$ and the period $\tau$, we increase the activation by letting $\ell$ grow (and $r$ in the same way).  The increase of $\ell$ (or $r$) at fixed $E_0$ and $\tau$ represents the amplitude of ``shaking'' of the field.

\begin{figure}[!t]
	\centering
	\includegraphics[width=7.2 cm]{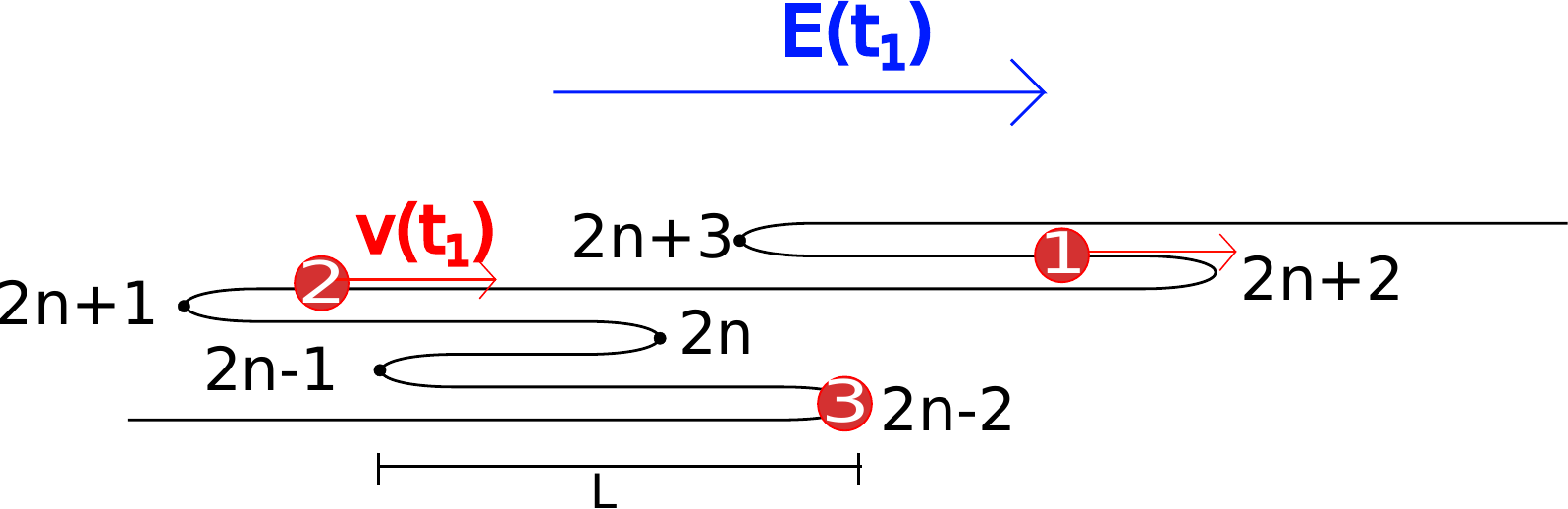}
	\includegraphics[width=7.2 cm]{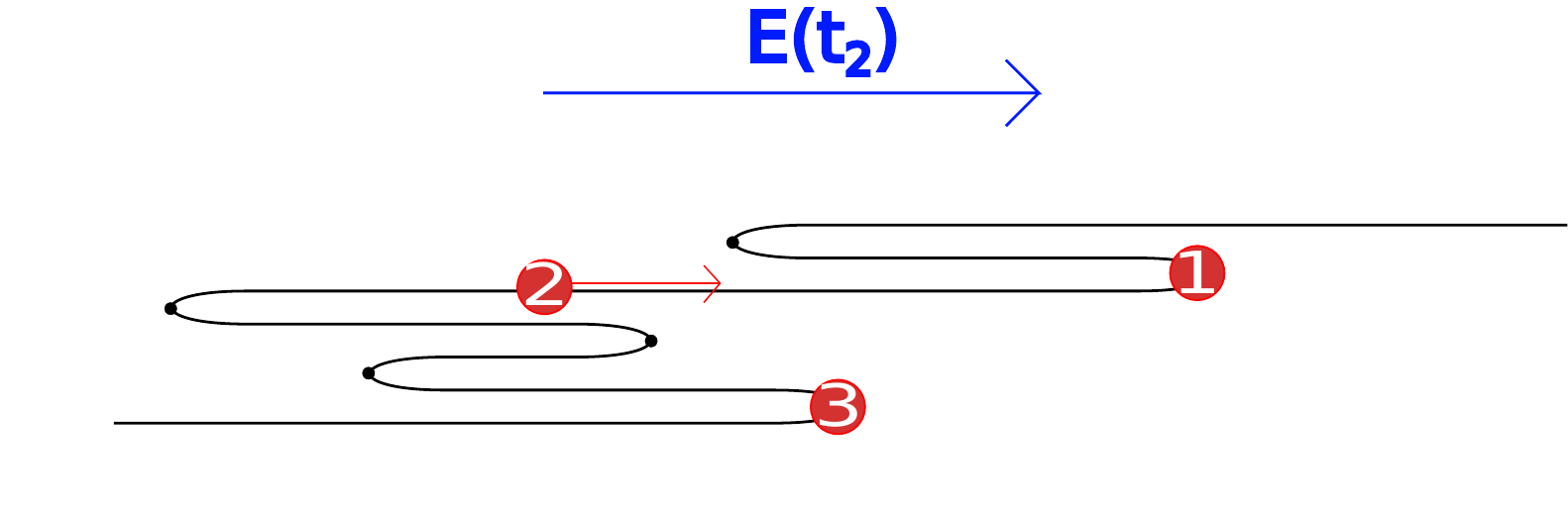}
	\caption{(Color online) Left: The initial condition for a system with three independent particles. See figure~\ref{field} to situate the time $t_1$. Right: At time $t_2$ particles 1 (always the one above) and 3 (always lowermost) get stuck in even-numbered nodes. They are staying there until the field reverses at time $t_3$. Particle 2 (always in the middle) currently continues its rectilinear course.} \label{conductivity}
\end{figure}

\begin{figure}[!t]
	\centering
	\includegraphics[width=7.2 cm]{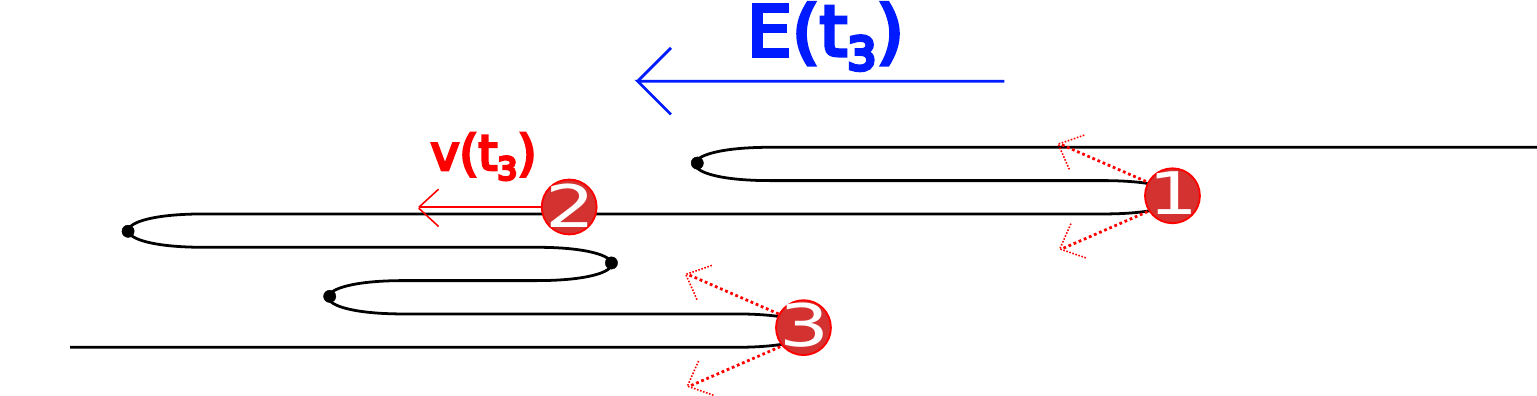}
	\includegraphics[width=7.2 cm]{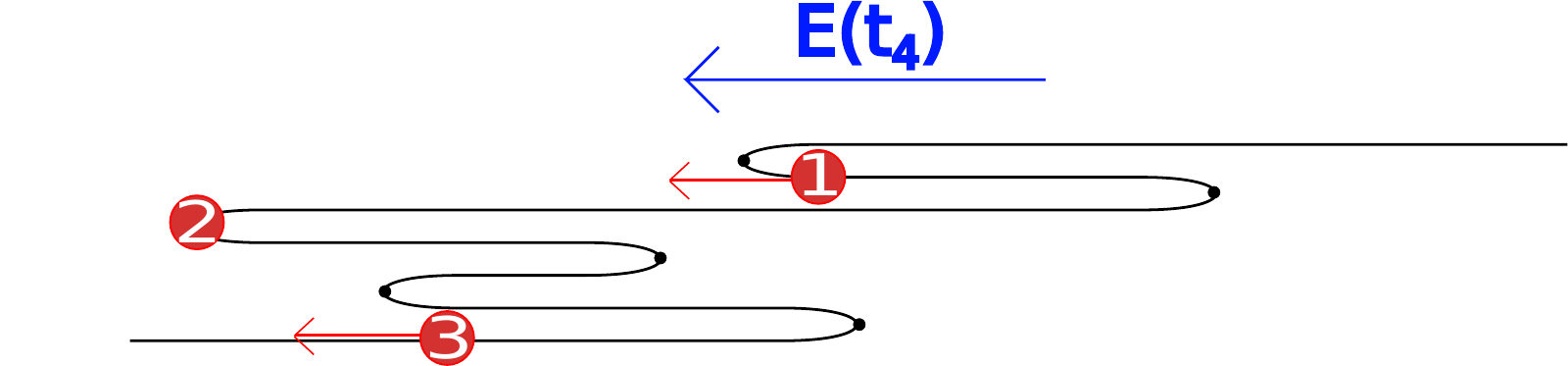}
	\caption{(Color online) Left: At time $t_3$, the field $E$ suddenly changes sign. Particles 1 and 3 immediately choose whether to follow the wire forward or backward. That choice is made at random and is independent of any other events in the dynamics. Immediately thereafter, all particles adopt the speed $v(t)=E(t)$. Right: At time $t_4$, particle 2 has gotten stuck in node $2n+1$, where it must remain until the next field-reversal at time $t_5$. We see that particle 1 has made the choice to go forward while particle 3 has gone backward: a scenario with probability 1/4 to proceed from the situation at time $t_3$.} \label{conductivity3}
\end{figure}

\begin{figure}[!t]
	\centering
	\includegraphics[width=7.2 cm]{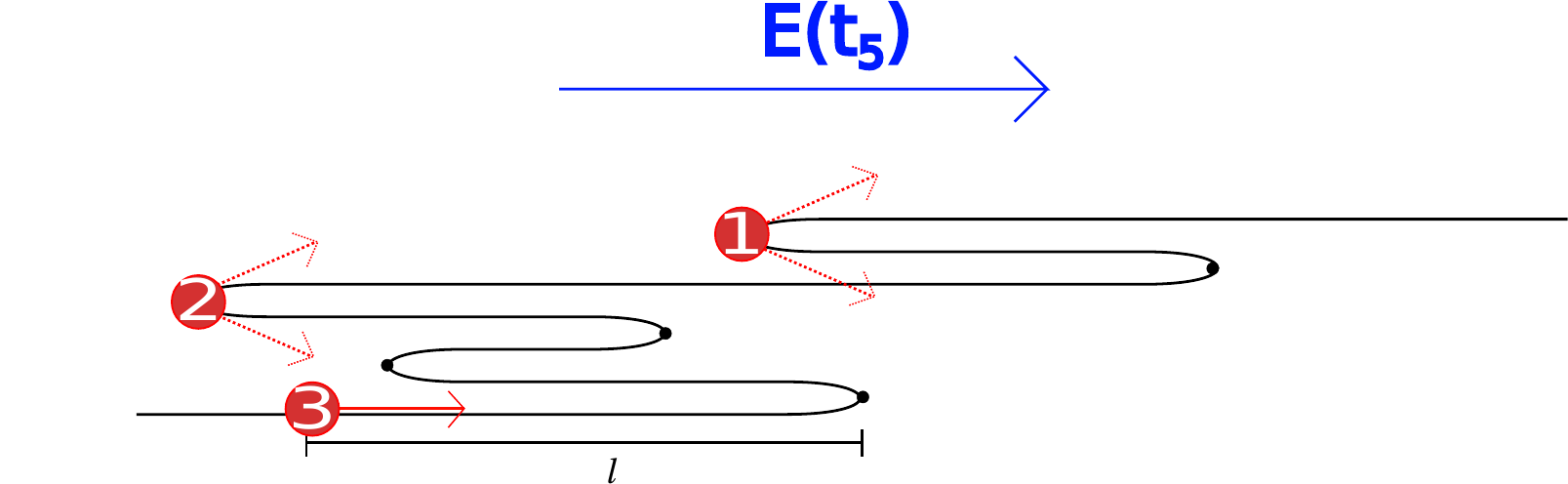}
	\includegraphics[width=7.2 cm]{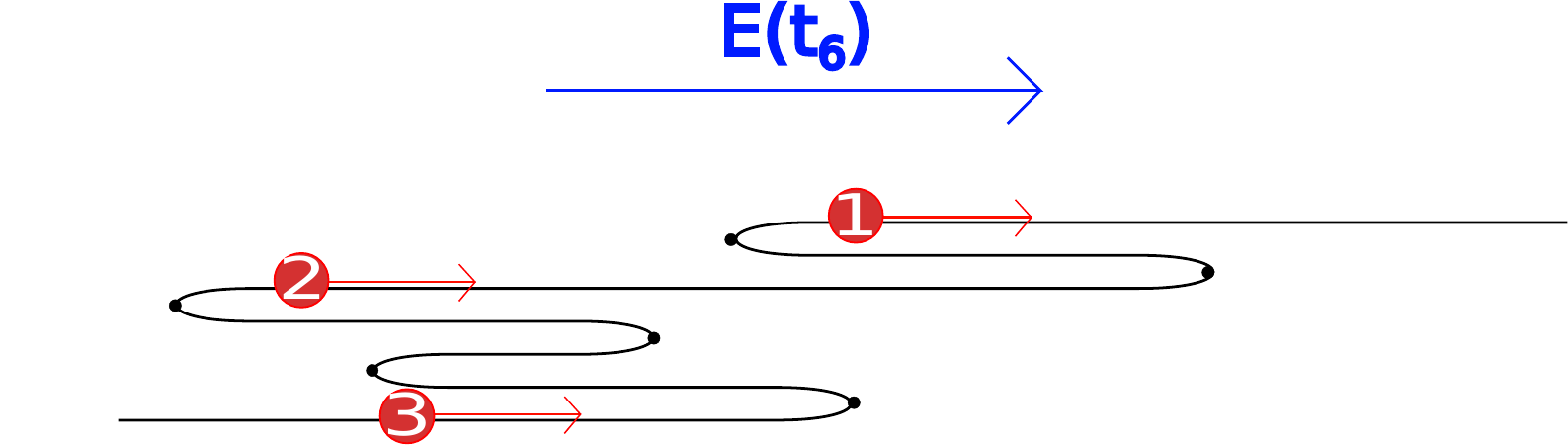}
	\caption{(Color online) Left: At time $t_5$, the field suddenly reverses once more. Particles 1 and 2 were previously stuck in odd nodes and, therefore, now make a random, unbiased choice to continue either forward or backward. Particle 3 has proceeded by a length $\ell$ --- i.e., the leftward tendency --- away from node $2n-2$. Right: A possible state of affairs at time $t_6$. There are 16 possible $t_6$-states for the initial condition at $t_1$, all of which occur with probability 1/16.} \label{conductivity5}
\end{figure}

In figures~\ref{conductivity}, \ref{conductivity3}, \ref{conductivity5} we play the dynamics following the times $t_i$ indicated in figure~\ref{field}.

There are three types of nodes in the random wire which merit special attention:
\begin{itemize}
\item A type-0 node is an even-numbered node ($2n$) which is inescapable. Particles which hit it at some point will thereafter never be capable of journeying to $2n-1$ or $2n+1$, let alone $2n-2$, $2n+2$. A system where type-0 nodes are in every positive and negative tail $\left(2n\right)_{n>M}$, $\left(-2n\right)_{n>M}$ does not allow for a nonzero current or diffusion as the particles must, almost surely, relax in a periodic orbit in the vicinity of a type-0 node.
\item A type-1 node is an even-numbered node ($2n$) where all journeys are possible. One can easily verify that within a single period $\tau$, a particle leaving that node will travel to $2n-2$ or $2n+2$ with probability 1/4 while the option to return to $2n$ has probability 1/2.
\item A type-2 node is an even-numbered node ($2n$) where only the journey to $2n+2$ is possible. Within a single period $\tau$  such a journey has probability 1/4, while returning to $2n$ happens with probability~3/4.
\end{itemize}

\subsection{First-order transition in the current}

We can write down an exact expression for the asymptotic time-averaged current $\bar{v}$ in a system where only nodes of type 1 and type 2 occur and are distributed according to a product measure, with $0\leqslant p\leqslant 1$ being the probability for a given node to be of type 1:
\begin{equation} \label{result}
\bar{v} = (1-p)\frac{p\ell_1+(1-p)\ell_2}{4\tau+(1-p)^2({\cal T}-\tau)}
\end{equation}
for almost all wire geometries.  The $\ell_1,\,\ell_2>0$ are disorder-averaged wire-lengths between a given node $2n-2$ and the node $2n$ conditioned on the latter being of type 1 and type 2, respectively. $\tau$ is the period of the field $E$ while ${\cal T}$ is the disorder-averaged journey-time for a particle to travel from a node $2n-2$ to the node $2n$, conditioned on the latter being of type 2. A derivation of formula \eqref{result} is given in the Appendix~\ref{A} accompanied by figure~\ref{1st}.
In formula \eqref{result} the bias $E_0$ is (only) in the ${\cal T}$.

\begin{figure}[!t]
	\centering
	\includegraphics[width=8 cm]{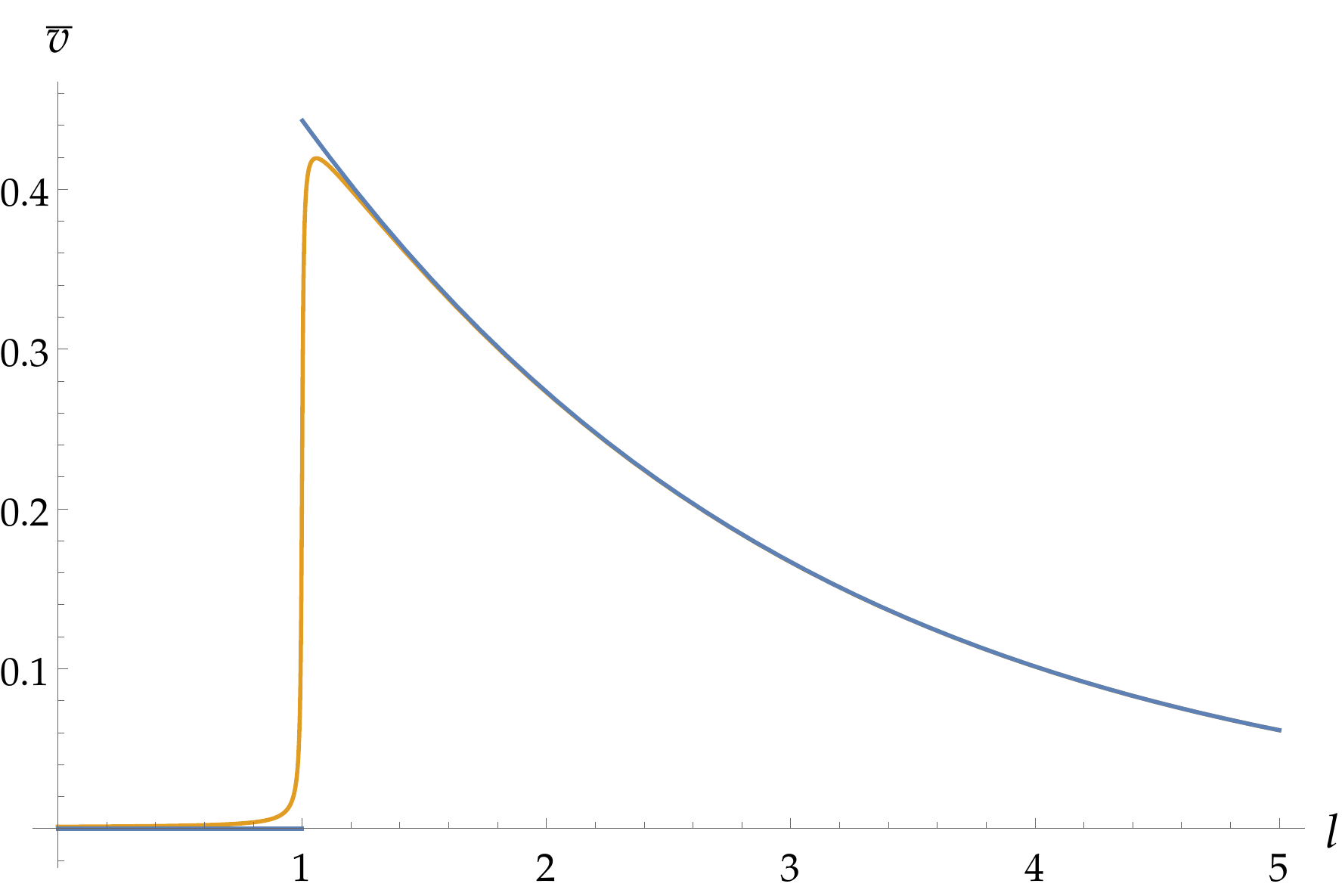}
	\caption{(Color online) The blue line represents the current for the wire-geometry discussed in the first example \eqref{1ex} with the values indicated under \eqref{1ex}. The orange line is the conjectured current if the temperature $T$ were small but nonzero and everything else is left unchanged.} \label{res}
\end{figure}

We make formula \eqref{result} explicit.
Imagine $L_-$ is exponentially distributed in length --- with average~$\lambda$ --- while $L_+=L$. As long as the activity of the field $\ell<L$ is strictly smaller than $L$, there are plenty of type-0 nodes  so that all particles enter localized periodic orbits and no current flows.
	However, as soon as $\ell \geqslant L$, the following values should be plugged into formula \eqref{result}:
	\begin{equation}\label{1ex}
	\begin{cases}
	& p = 1-\exp(-\ell/\lambda), \\
	& \ell_1=L+\lambda-\ell\frac{\exp\left(-\ell/\lambda \right)}{1-\exp\left(-\ell/\lambda \right)}\,, \\
	& \ell_2=L+\ell+\lambda, \\
	&{\cal T}=\tau\left\{1+\frac{\exp[-(r-\ell)/\lambda]}{1-\exp[-(r-\ell)/\lambda]}\right\}.
	\end{cases}
	\end{equation}
	The resulting current for $L=1$, $\tau=1$, $\lambda=2$ and $r-\ell=3$ is given in figure~\ref{res} (blue curve).  The two regimes are summarized in figure~\ref{1st}.  Note that this scenario is strictly at zero-temperature.  We expect the orange curve in figure~\ref{res} to represent the smoothened-out resurrection at low temperature. Finally, for this model of the geometry of the wire, we can calculate the linear response regime when $E_0=(r-\ell)/\tau$ is small.  In that linear regime, we, of course, have zero current, $\bar{v} = 0$, when $\ell < L$ and otherwise, by expansion in $E_0$ for fixed activation
	$\ell$,
	\[
	\bar{v} = \frac{p \ell_1+(1-p) \ell_2}{(1-p) \lambda} E_0.
	\]

\begin{figure}[!b]
\vspace{-6mm}
	\centering
	\includegraphics[width=10 cm]{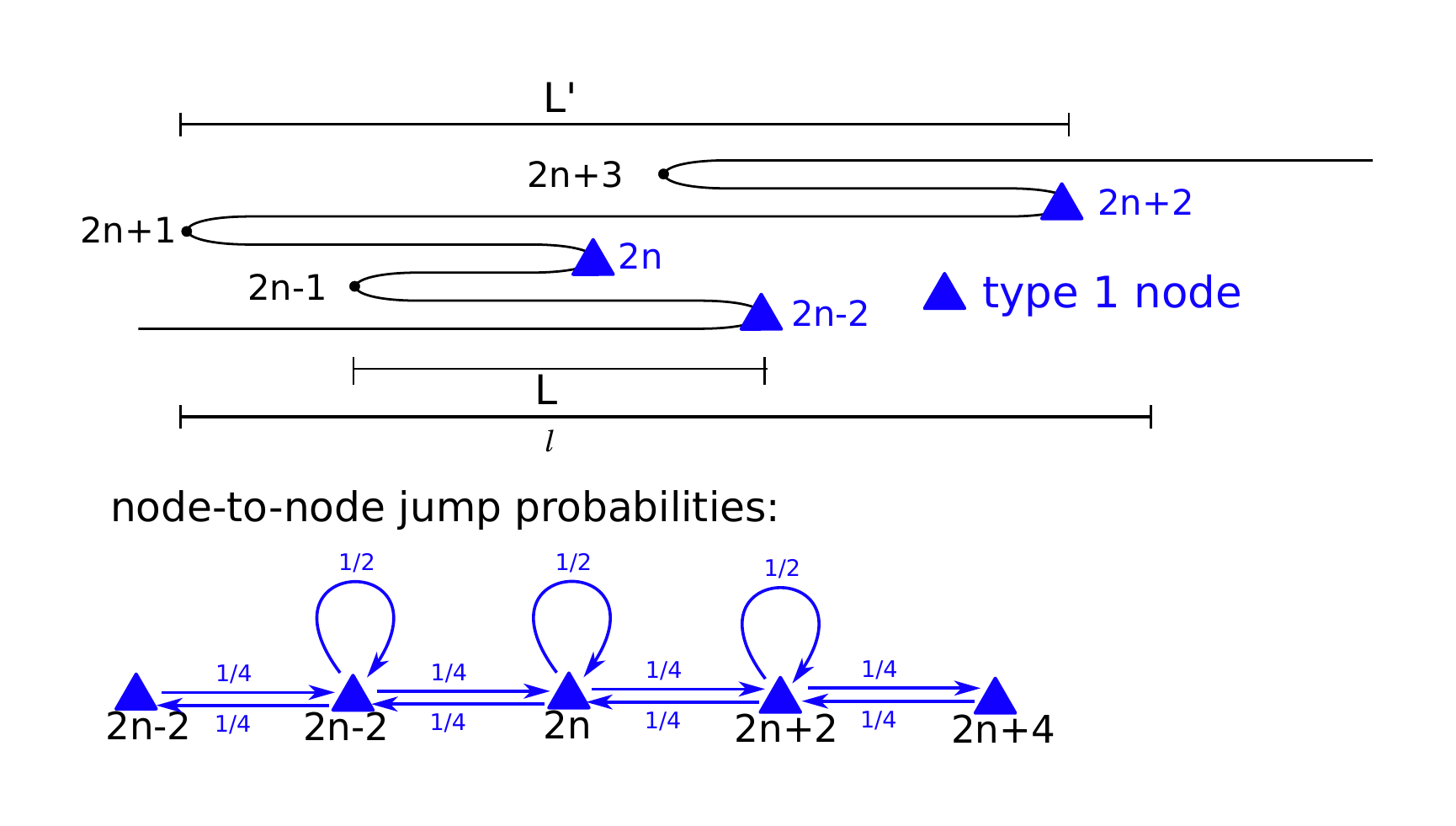}
	\caption{(Color online) A possible super-active phase: such a regime occurs when there is also a cut-off $L'$ on the length of rightward-directed strands. When the leftward-tendency $\ell\geqslant L'$, one has a regime of activity where particles can diffuse unhindered along the chain (due to the external field). Yet, the external bias of the field is no more translated into a current/ballistic motion.} \label{superactive}
\end{figure}

\begin{figure}[!b]
	\centering
	\includegraphics[width=8 cm]{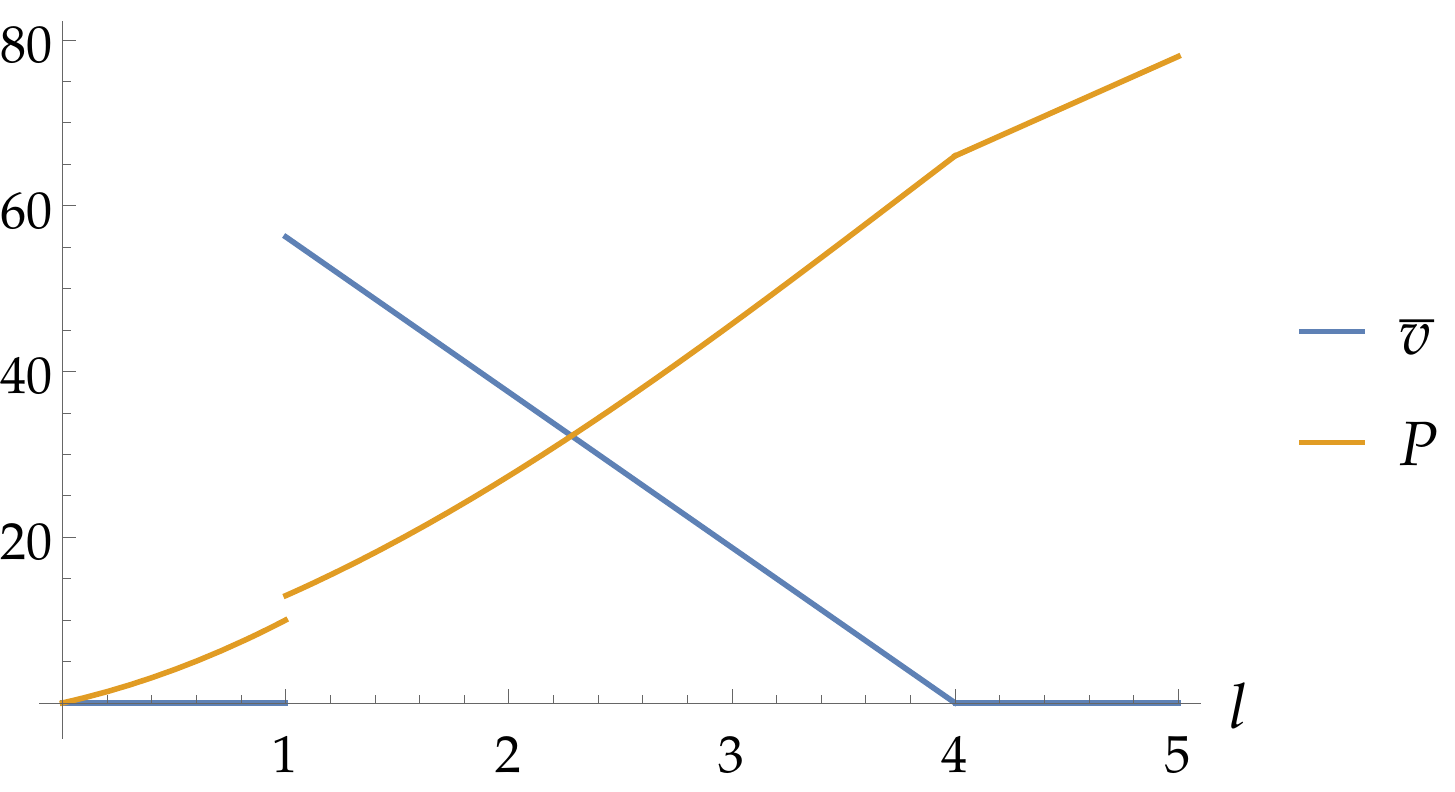}
	\caption{(Color online) The current $\bar{v}$ and dissipation $P$ --- or mean entropy production rate --- in the system discussed in the second example as a function of the activation $\ell$. Again for $L=1$, $L'=4$, $\tau=1$ and $r-\ell=3$. The calculation of the dissipation is skipped here.} \label{diss}
\end{figure}

Consider a second example: suppose now that $L_-$ is homogeneously distributed in the interval $[0,L']$ while still $L_+=L$. Again, no current flows as long as $\ell<L'$.
	However, when $L\leqslant \ell \leqslant L'$, the following values have to be plugged into formula \eqref{result}:
	\begin{equation}\label{ex2}
	\begin{cases}
	& p = \frac{\ell}{L'}\,, \\
	& \ell_1=L+\frac{\ell}{2}\,, \\
	& \ell_2=L+\frac{L'+\ell}{2}\,, \\
	& {\cal T}=\tau \qquad\text{ (at least when $L=1$, $L'=4$, $\tau=1$, and $r-\ell=3$)}.
	\end{cases}
	\end{equation}
	 The current corresponding to those values $L=1$, $L'=4\,,\ldots$ is again given in figure~\ref{res} (green curve). For $\ell>L'$, the current is again zero, while, on the other hand, the diffusion of the particle has not lessened; see figure~\ref{superactive}. In figure~\ref{diss}, we show the dissipated work for the same system.

\section{Conclusion: on the role of activation}

Activation in nonequilibrium situations as in the above has been less often considered.  Yet, it appears in many kinetic questions, such as in the problem of relaxation to equilibrium.  The reason why systems do not relax instantaneously is, of course, related to the fact that the metric for steepest descent in a free energy landscape is controlled by factors of activity.
Let us take an example of a Markov jump process to illustrate the issue more concretely.
To make it very simple, we consider a mesoscopic process where the reduced system has states $x = 1,2,\ldots,n$, thought here as the positions on a ring of size $n$. Each $x$ thus represents a state of an $N$-particle system. On the considered level of description, we assume a Markov evolution for a random walker with transition rates
\[
k(x,x+1) = a_x\exp\{-\beta/2\left[ V(x+1) - V(x)\right]\}, \qquad k(x,x-1) = a_{x-1}\exp\{-\beta/2\left[ V(x-1) - V(x)\right]\}.
\]
Note that to each pair $\{x,x+1\}$ we have associated an activity or frequency $a_x>0$, telling us the local time-scale for the traffic over $x\leftrightarrow x+1$.  Other choices and parameterizations are possible and sometimes wished in the case of Arrhenius kinetics but the main idea of what follows does not depend on it.
For every choice, though there is a detailed balance for the potential $V$ at inverse temperature $\beta$,
\[
\frac{k(x,x+1)}{k(x+1,x)} = \exp\{-\beta\left[ V(x+1) - V(x)\right]\}
\]
and indeed the Boltzmann-Gibbs weight $\sim \exp[-\beta V(x)]$ gives the stationary occupation at site $x$.  The equation describing relaxation to equilibrium is here the Master equation, but let us simply imagine that at some time all particles sit at position $x$.  There are two interesting questions now concerning the relaxation; one is what the preferred next position is, and the other is about the time it takes to make the move. For the first question, we need to know the fraction of particles that go right $x\rightarrow x+1$ to those that go left $x\rightarrow x-1$.  Clearly, that is given by the ratio of fluxes
\begin{equation}\label{dp}
\frac{ a_x\exp[\beta V(x-1)/2]}{a_{x-1}\exp[\beta V(x+1)/2]}.
\end{equation}
Hence, when $a_x \gg a_{x-1}$ particles will typically flow to the right when the potential does not vary much, $V(x-1)\simeq V(x+1)$. Even when $ V(x-1)\leqslant V(x+1)$ and the potential would guide the particles to the left, still the reactivities decide when \eqref{dp} is very large.  That is a kinetic effect, not visible from the stationary occupations which are determined thermodynamically.  In the long run, the effect cancels, and the occupations are determined by the potential and the temperature and all currents disappear.  The important reminder is that time-symmetric effects as quantified here in the local frequencies $a_x$ can determine the current direction in the transient regime for the relaxation to equilibrium and they are, of course, directly giving a measure of the dynamical activity over the corresponding bonds $x \leftrightarrow x+1$; see more in \cite{mos,heatb}.

For the second question, we take all $a_x = a$ fixed while we assume that for $x$, the differences $\beta[V(x+1) - V(x)]$ and $\beta[V(x-1) - V(x)]$ are so big that effectively the particles prefer to stay in position~$x$.  That is to say, that the escape rate
$k(x,x+1) + k(x,x-1) \sim a\exp(-N)$ is minute for some big~$N$, where $N$ could be an extensive parameter indeed as the number of particles or scaling like the volume of the total system. Remember that we are dealing here with an effective and coarse-grained system where we assume that $x$ is such that an escape from it is thermodynamically not available.  The system is then essentially trapped in condition $x$ and will be seen to reside there for macroscopic times.  However, if we are able to change the activities $a \rightarrow a^{sN}$ in the same way for some parameter $s>0$, we will again see the activity and eventual relaxation to equilibrium.  The reason is clear from looking at the escape rates, but there is a more interesting calculation which involves the path-probabilities and which makes contact with spatially extended models.  The point is that the probability of a trajectory $\omega = (x_s, 0\leqslant s\leqslant t)$ scales like
\[
\text{Prob}[\omega] \propto \prod_x\left( a^{{\cal K}_{x,x+1}^t}\right)\,\exp\Bigg[-\int_0^t \id \tau\,\xi(x_\tau)\Bigg]\exp\left\{\frac{\beta}{2}[V(x_0)-V(x_t)]\right\},
\]
where ${\cal K}_{x,x+1}^t$ is the number of jumps over $x\leftrightarrow x+1$ (always counted positive) in the time-interval $[0,t]$.
Trajectories where $\beta[V(x_t)-V(x_0)] \sim N \gg 1$ are very damped, unless we change the $a\rightarrow a^{sN}$, which amounts here to changing the dynamical ensemble into
\[
\text{Prob}_s[\omega] = \text{Prob}[\omega] \exp \Big(sN \log a\sum_x {\cal K}_{x,x+1}^t\Big).
\]
We have ignored the change in the escape rates $\xi$'s for which we assume that they remain of the order one.  We note that for $s<0$, the system would not show any activity or relaxational behaviour, while for $s>0$, the system is biased to show the activity and would relax to equilibrium.  That is not unlike the dynamical phase transitions discussed in \cite{dpt,kin,chan}.

We finally remark that the opposite scenario is much better known.  Suppose indeed that $a \simeq a_o\exp(-\ell)$ where $\ell$ can be huge depending on certain parameters or constraints in the total system.  Even when the escape from $x$ is thermodynamically allowed, we need activation to overcome the barrier set by the $a$'s.

The point of the paper has been to illustrate the previous formalities for a more specific set-up, where (1) the current in a steady system dies when the amplitude of a forcing exceeds a certain value, and (2) the current resurrects to a non-zero saturation value when the forcing is appropriately modulated in time.  The latter, the time-modulation, is thus a physical way of introducing activation.  It effectively increases the activities and results in the flowing of a current which previously vanished.

\section*{Acknowledgement}

We thank Pieter Baerts for help in providing the figures, and we are grateful to Luca Avena, Frank den Hollander, Fran\c cois Huveneers and Thimoth\'ee Thiery for private communication and discussions.

\appendix
\section{Derivation of current formul{\ae} for the current} \label{A}

\begin{figure}[!b]
		\centering
		\includegraphics[width=7.2 cm]{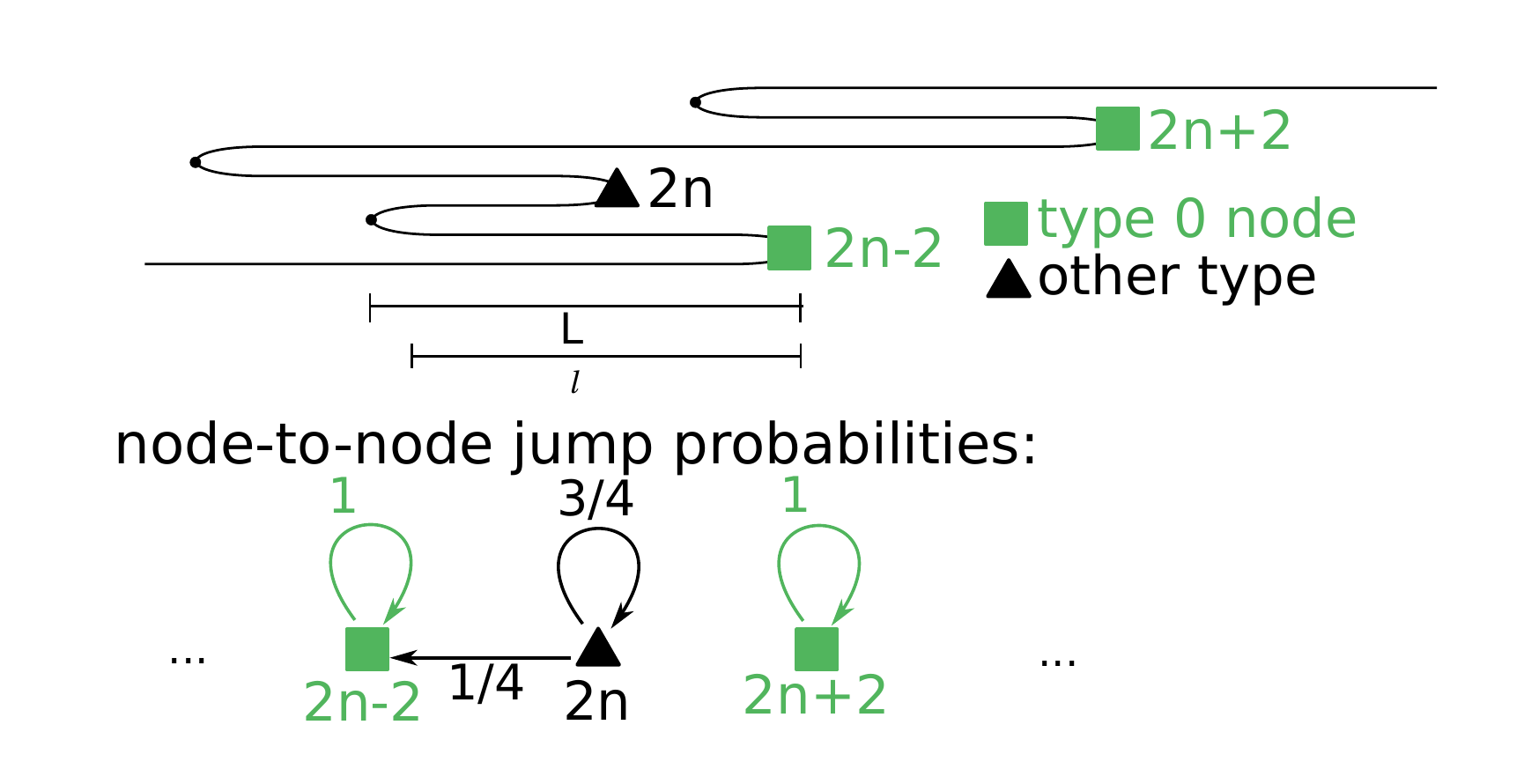}
		\includegraphics[width=7.2 cm]{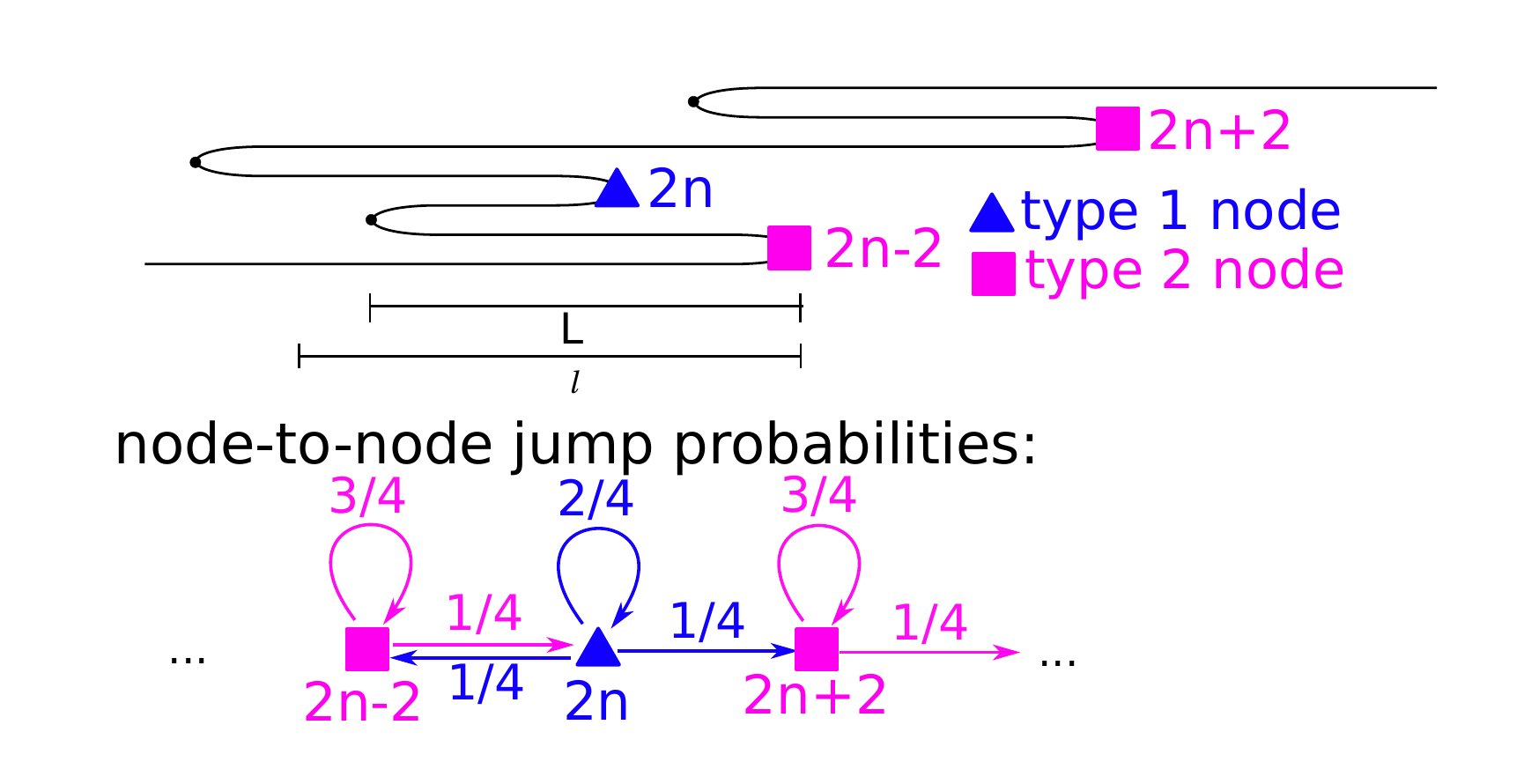}
		\caption{(Color online) Left: The inactive phase, $\ell<L$. Particles do not flow nor diffuse along the chain. After a sufficient time, they settle down into a periodic, localized orbit situated in the vicinity of a single even node of type 2. Right: The active phase, $\ell\geqslant L$. Particles suddenly start flowing, giving rise to a nonzero current. Though nodes of type 2 prevent orbits deviating to $-\infty$, the motion is nonetheless also diffusive in the sense that $\text{Var}(x(t))\propto t$ for large $t$.}\label{1st}
	\end{figure}

We refer to the figure~\ref{1st}.
	
Suppose the activation $\ell>L$, so that only nodes of type 1 and type 2 remain. 
We partition the wire into cells $(c_{k})_{k\in \mathbb{Z}}$ which begin and end in a node of type 2 with only nodes of type 1 in between. Instead of constructing the wire by choosing the connecting strands as the random variables $L_-$ and $L_+$, one can devise a scheme where the wire is constructed by choosing the cells i.i.d. from an appropriate probability space $C$. Any element $c \in C$ has a random number of type 1 nodes included: the probability for precisely $m$ such nodes is given by $p^m(1-p)$. Moreover, the element $c$ also carries the information on the strand-lengths connecting all its constituent nodes. In particular, we denote by $|c|$ the sum of all $c$'s strand-lengths.

Next, we create another auxiliary probability space $X$. The elements $\Xi=(c,o)\in X$ are doublets where $c \in C$ is a cell as described before. $o=(n_1(o),n_2(o),n_3(o),t(o))\in \mathbb{N}^3\times \mathbb{R}$ is the essential information of an orbit-fragment obtained by dropping a particle at the left-end of the cell $c$, starting the external field $E(t)$ at time $t=0$ and playing the dynamics just until the (first) time $t(o)=t(\Xi)\in \tau \mathbb{N}$ where the particle has arrived at the right-end of the cell $c$ and the field has just completed a period. $n_1(o)$ is the orbit $o$'s total number of journeys from the leftmost (type 2) node to itself. Similarly, $n_2(o)$ is the total number of journeys starting at a type 1 node and ending at the same node. $n_3(o)$ is the number of journeys where the destination differs from the starting point. Suppose $Q\subset C$ is a set of cells such that $\forall c \in Q$, $o=(n_1,n_2,n_3,t)$ is a possible orbit. Then, the probability measure $\text{d} \mu$ on $X$ is defined by $\text{d} \mu \left(Q\times \left\{o\right\}\right)=\text{d} \mu_c(Q)\left(\frac{3}{4}\right)^{n_1}\left(\frac{1}{2}\right)^{n_2}\left(\frac{1}{4}\right)^{n_3}$ where $\text{d} \mu_c$ is the probability measure inherited from the space~$C$.

 We carry over the notation $|\Xi|=|(c,o)|=|c|$ to denote the total cell-length.

\subsection{The expected cell-length}
The expectation value for $|\Xi|=|c|$ is given by
\begin{align}
\mathbb{E}_X[|\Xi|]&=\sum_{j=0}^{+\infty} p^j(1-p) \left\{(j+1)\mathbb{E}[L_+]+j\mathbb{E}[L_-\left.\right|L_- \leqslant \ell]+\mathbb{E}[L_-\left.\right|L_- > \ell]\right\} \nonumber\\
&=\sum_{j=0}^{+\infty} p^j(1-p) \left(j\ell_1+\ell_2\right)  = \ell_1\frac{p}{1-p}+\ell_2.
\end{align}

\subsection{The expected time to traverse a cell}
\begin{equation}\label{exp}
\mathbb{E}_X[t(\Xi)]=\sum_{j=0}^{+\infty} p^j(1-p) \underbrace{\mathbb{E}_X[t(\Xi)\left.\right| \Xi=(c,o) \text{ has }j \text{ nodes of type 1}]}_{=:\tau(j)}.
\end{equation}
Now, for every journey which ends in a type 1 node, the travel time is a single period $\tau$. The same is true for journeys from a type 2 node to itself. So, an orbit $o\leftrightarrow \Xi=(o,c)$ which needs $J(\Xi)\in\mathbb{N}$ jumps to traverse the cell $c$ will last for a duration of $[J(\Xi)-1]\tau+\hat{\cal{T}}(\Xi)$. (The last jump was from a node of type 1 toward a node of type 2 of the next cell and, therefore, lasts a time $\hat{\cal{T}}(\Xi)=\hat{\cal{T}}(c) \in \tau \mathbb{N}$ which is random but only dependent on the strand-length between the two final nodes of the cell. We denote by $\cal{T}$ the expectation value of the latter variable.) So, calculating $\tau(j)$ is equivalent to calculating
\begin{equation}
\mathbb{E}_X[J(\Xi)\left.\right| c \leftrightarrow \Xi=(c,o) \text{ has }j \text{ nodes of type 1}].\nonumber
\end{equation}
However, this can be done: in a cell with $j$ intermediate nodes, the internode-journeys are governed by the following transition-matrix (which acts on a column-vector of site-occupation probabilities once every period $\tau$):
\begin{equation}
P = \text{diag}\Bigg(\frac{3}{4}, \underbrace{\frac{1}{2},\ldots,\frac{1}{2}}_{\times j}\Bigg) + \frac{1}{4}(S+S^{\text{T}}),\nonumber
\end{equation}
where $S$ is the elementary nilpotent $(j+1)\times (j+1)$ matrix
\begin{equation}
S = \begin{pmatrix}
& 0 & 1 & 0 & \ldots&\\
& 0  & 0 & 1 & \ldots &\\
& \vdots & \vdots & &\vdots&\vdots \\
& & &\ldots & 0 & 1\\
& & & \ldots& 0 & 0
\end{pmatrix}.\nonumber
\end{equation}
Notice that the entries of the last column of $P$ sum only to $3/4$: with probability $1/4$, a particle at the last site will jump to the next cell. Initially, the particle starts at the first site, hence $\rho(0)=(1,0,\ldots,0)^\text{T}$. Then, the future probabilities are given by $\rho(J)=P^J\rho(0)$. The proportion of orbits, where the particle leaves precisely after $J$ periods, is given by $\sum_{m=1}^{j+1}[\rho(J-1)]_m-[\rho(J)]_m = \sum_{m=1}^{j+1} [(P^{J-1}-P^{J})\rho(0)]_m$. So, we conclude that
\begin{align*}
&\mathbb{E}_X[J(\Xi)\left.\right| \Xi \text{ has }j \text{ nodes of type 1}] = \sum_{J=1}^{+\infty} J \sum_{m=1}^{j+1} [(P^{J-1}-P^{J})\rho(0)]_m=\sum_{m=1}^{j+1}\sum_{J=0}^{+\infty}  [P^{J}\rho(0)]_m \\
& = \sum_{m=1}^{j+1}  [(1-P)^{-1}\rho(0)]_m  =\sum_{m=1}^{j+1} (j+2-m)4 = 4 \frac{(j+1)(j+2)}{2}=2(j+1)(j+2),
\end{align*}
where we computed the entries of the vector $(1-P)^{-1}\rho(0)$, i.e., the first column of $(1-P)^{-1}$, explicitly.
Plugging this result into $\tau(j)=\langle J-1 \rangle \tau + \cal{T}=2(j+1)(j+2)\tau + (\cal{T}-\tau)$. Plugging this into \eqref{exp} and performing the sums yields
\begin{equation}
\mathbb{E}_X[t(\Xi)]=\frac{4\tau}{(1-p)^2}+\cal{T}-\tau.
\end{equation}

\subsection{The current}
Let $(\Xi_k)_{k\in\mathbb{Z}}=(c_k,o_k)_{k\in\mathbb{Z}}$ be an i.d.d. sequence of cells $c_k$ and the orbit-fragment $o_k$ therein. The average velocity $v_n$ after $n$ cells  is given by
\begin{equation}
v_n=\frac{|\Xi_1|+\ldots+|\Xi_n|}{t(\Xi_1)+\ldots+t(\Xi_n)} = \frac{\frac{1}{n}\left(|\Xi_1|+\ldots+|\Xi_n|\right)}{\frac{1}{n}\left[t(\Xi_1)+\ldots+t(\Xi_n)\right]}\,.\nonumber
\end{equation}
However, in that expression, the law of large numbers implies convergence (respectively in probability, or strictly almost surely) for $n \to \infty$ in both the numerator and the denominator. So, defining $\bar{v}=\lim_{n \to \infty} v_n$, we get
\begin{equation}
\bar{v}=\frac{\mathbb{E}_X[|\Xi|]}{\mathbb{E}_X[t(\Xi)]}=\frac{\ell_1\frac{p}{1-p}+\ell_2}{\frac{4\tau}{(1-p)^2}+\cal{T}-\tau}=(1-p)\frac{p\ell_1+(1-p)\ell_2}{4\tau+(1-p)^2(\cal{T}-\tau)}\,.
\end{equation}
It is not hard to show that the related current $\bar{v}'=\lim_{t \to \infty} \frac{x(t)}{t}$ is almost surely equal to $\bar{v}$ since the orbit is at every time situated between the right-ens of some cell $k$ and $k+1$ and, therefore,
\begin{equation}
\frac{|\Xi_1|+\ldots+|\Xi_{k}|}{t(\Xi_1)+\ldots+t(\Xi_{k+1})}\leqslant \frac{x(t)}{t} \leqslant \frac{|\Xi_1|+\ldots+|\Xi_{k+1}|}{t(\Xi_1)+\ldots+t(\Xi_k)}\nonumber
\end{equation}
from where the equality \eqref{result} follows by sandwiching.

\ukrainianpart

\title{Індукований активністю перехід першого роду для струму в невпорядкованому середовищі}
\author{T. Демаєрель, К. Маєс}
\address{Інститут теоретичної фізики, Льовенський католицький університет, Бельгія}

\makeukrtitle

\begin{abstract}
Відомо, що частинки можуть попасти в пастку завдяки випадково розташованим перешкодам, якщо вони зазнають сильних зіткнень.
Ми представляємо модель, де струм у невпорядкованому середовищі зникає при великому зовнішньому полі, але відновлюється при зростанні активності. Під активністю ми маємо на увазі часову варіацію зовнішнього усередненого за часом постійного поля. Інше пояснення відновлення струму полягає в тому, що частинки здатні проходити нескінченну послідовність потенціальних бар'єрів через механізм,  подібний до стохастичного резонансу. Ми також обговорюємо роль ``струшування'' в процесах релаксації.
\keywords випадкове середовище, нерівноважний перехід першого роду
\end{abstract}


\begin{thebibliography}{10}

\bibitem{snit}
 Sznitman A.-S., Brownian Motion, Obstacles and Random Media, Springer-Verlag, Berlin, 1998.


\bibitem{Barma}
 Barma M., Dhar D., J. Phys. C: Solid State Phys., 1983, {\bf 16}, 1451, \doi{10.1088/0022-3719/16/8/014}. 

\bibitem{fre}
 Leitmann S., Franosch T., Phys. Rev. Lett., 2013, {\bf 111}, 190603, \doi{10.1103/PhysRevLett.111.190603}.

\bibitem{ben}
 B\'enichou O., Illien P., Oshanin G., Sarracino A., Voituriez R.,
Phys. Rev. Lett., 2014, {\bf 113}, 268002, \doi{10.1103/PhysRevLett.113.268002}.

\bibitem{tra}
 Ramaswamy R., Barma M., J. Phys. A: Math. Gen., 1987, {\bf 20}, 2973, \doi{10.1088/0305-4470/20/10/039}. 

\bibitem{bae}
 Baerts P., Basu U., Maes C., Safaverdi S., Phys. Rev. E, 2013, {\bf 88}, 052109, \doi{10.1103/PhysRevE.88.052109}.
 
\bibitem{kolk}
 Basu U., Maes  C., J. Phys. Conf. Ser., 2015, {\bf 638}, 012001, \doi{10.1088/1742-6596/638/1/012001}.

\bibitem{gar}
 Everest B., Lesanovsky I., Garrahan J.P., Levi E., Phys. Rev. B, 2017, {\bf 95}, 024310,\\ \doi{10.1103/PhysRevB.95.024310}.

\bibitem{dpt}
 Garrahan J.P., Jack R.L., Lecomte V., Pitard E., van Duijvendijk K., van Wijland F.,
 J. Phys. A: Math. Theor., 2009, {\bf 42}, 075007, \doi{10.1088/1751-8113/42/7/075007}.

\bibitem{kin}
 Garrahan J.P., Sollich P., Toninelli C., In: Dynamical Heterogeneities in Glasses, Colloids, and Granular Media, Berthier L., Biroli G., Bouchaud J.-P., Cipelletti L., van Saarloos W.~(Eds.), Oxford University Press, Oxford, 2011, 341--369, [Preprint \arxiv{1009.6113}, 2010].

\bibitem{chan}
 Jack R.L., Garrahan J.P., Chandler D., J. Chem. Phys., 2006, {\bf 125}, 184509, \doi{10.1063/1.2374885}.

\bibitem{mob}
 Basu U., Maes C., J. Phys. A: Math. Theor., 2014, {\bf 47}, 255003, \doi{10.1088/1751-8113/47/25/255003}.

\bibitem{Zia}
 Zia R.K.P.,  Praestgaard E.L., Mouritsen O.G., Am. J. Phys., 2002, {\bf 70}, 384, \doi{10.1119/1.1427088}. 

\bibitem{sol}
 Solomon F., Ann. Probab., 1975, {\bf 3}, 1--31, \doi{10.1214/aop/1176996444}. 

\bibitem{scholar}
 Rouvas-Nicolis C., Nicolis G., Scholarpedia, 2007, \textbf{2}, No. 11, 1474, \doi{10.4249/scholarpedia.1474}.

\bibitem{mos}
 Maes C.,  Math. Mech. Complex Syst., 2016, \textbf{4}, 275--295, \doi{10.2140/memocs.2016.4.275}.

\bibitem{heatb}
 Maes C., Neto\v cn\'y K., Ann. Henri Poincar\'e, 2013, {\bf 14}, 1193--1202, \doi{10.1007/s00023-012-0214-8}.

\end{thebibliography}
\end{document}